\documentclass[preprint,12pt,authoryear]{elsarticle}
\usepackage{amssymb,latexsym,amsthm,amsmath,graphicx,color,hyperref,url}
\usepackage[ruled,vlined]{algorithm2e}

\journal{Theoretical Computer Science}

\newtheorem{theorem}{Theorem}

\newtheorem{example}[theorem]{Example}
\newdefinition{definition}{Definition}
\newdefinition{remark}{Remark}
\newproof{pf}{Proof}

\newcommand{\ie}{\emph{i.e.,}}

\newcommand{\eg}{\emph{e.g., }}

\begin{document}

\begin{frontmatter}

\title{Parameter estimation for Boolean models of biological networks}

\author[clemson]{Elena Dimitrova}
\ead{edimit@clemson.edu}

\author[shsu,samsi]{Luis David Garc{\'\i}a-Puente}
\ead{lgarcia@shsu.edu}

\author[vt,vbi]{Franziska Hinkelmann}
\ead{fhinkel@vt.edu}

\author[vt,vbi]{Abdul S. Jarrah}
\ead{ajarrah@vbi.vt.edu}

\author[vt,vbi,samsi]{Reinhard Laubenbacher\corref{cor1}}
\ead{reinhard@vbi.vt.edu}

\author[smu,samsi]{Brandilyn Stigler}
\ead{bstigler@smu.edu}

\author[cornell]{Michael Stillman}
\ead{mike@math.cornell.edu}

\author[dimacs]{Paola Vera-Licona}
\ead{mveralic@math.rutgers.edu}

\cortext[cor1]{Corresponding author}
\fntext[fn1]{Partially supported by SAMSI New Researcher Fellowship.}

\address[clemson]{Department of Mathematical Sciences,
Clemson University, Clemson, SC  29634-0975, USA}

\address[shsu]{Department of Mathematics and Statistics,
Sam Houston State University, Huntsville, TX  77341-2206, USA}

\address[vt]{Department of Mathematics,
Virginia Polytechnic Institute and State University, Blacksburg, VA 24061-0123, USA}

\address[vbi]{Virginia Bioinformatics Institute,
Virginia Polytechnic Institute and State University, Blacksburg, VA 24061-0477, USA}

\address[smu]{Mathematics Department,
Southern Methodist University, Dallas, TX 75275-0156, USA}

\address[dimacs]
{DIMACS Center, Rutgers University,
Piscataway, NJ 08854-8018, USA}

\address[cornell] {Mathematics Department, Cornell University, Ithaca, NY 14853-4201, USA}

\address[samsi]
{Statistical and Applied Mathematical Sciences Institute,
Research Triangle Park, NC 27709-4006, USA}

\date{\today}

\begin{abstract}
Boolean networks have long been used as models of molecular networks and play an
increasingly important role in systems biology. This paper describes a software package, \emph{Polynome},
offered as a web service, that helps users construct Boolean network models based on
experimental data and biological input. The key feature is a discrete analog of parameter
estimation for continuous models. With only experimental data as input, the software
can be used as a tool for reverse-engineering of Boolean network models from experimental
time course data. 
\end{abstract}


\end{frontmatter}

\section{Introduction}


During the last decade finite dynamical systems, that is, discrete dynamical
systems with a finite phase space, have been
used increasingly in systems biology
to model a variety of biochemical networks, such as metabolic, gene regulatory, and
signal transduction networks.  In many cases, the available data quantity and quality
is not sufficient to build detailed quantitative models such as systems of ordinary
differential equations, which require many parameters that are frequently unknown.
In addition, discrete models tend to be more intuitive and more easily accessible
to life scientists. Boolean networks and the more general
so-called logical models are the main types of finite dynamical
systems that have been used successfully in modeling biological networks.

Discrete dynamical models of biological networks were first introduced by
Kauffman who used Boolean networks to study the dynamics of gene regulatory networks \citep{Kauffman1:69,Kauffman2:69,Kauffman:1993}.
A gene is assumed to be in one of two states, \emph{expressed} ($1$) or \emph{not expressed} ($0$).
The next state of a gene is determined by a Boolean function
in terms of the current states of the gene and its immediate neighbors in the network.
The state of a network in $n$ variables is then a binary vector of length $n$,
representing the state of each node of the network. Thus,  there are $2^n$ possible states. The
dynamics of the network is represented by a directed graph on the $2^n$ states, where each state has
out-degree one, that is, each state is mapped to exactly one other state (possibly itself).

Boolean  models of biological systems are abundant, including
gene regulatory networks such as
the segment polarity network in the fruit fly \citep{AO},
the cell cycle in mammalian cells \citep{AdrienFaure}, in budding yeast \citep{Li_cc_2004},
and fission yeast \citep{Bornhold_cc_plos}, and metabolic networks in \emph{E. coli}
\citep{jain_sysBio_2008,Ecoli_BN} and in \textit{S. cerevisiae} \citep{S.cerevisiae_BN}.
Also, Boolean network models of signaling networks have recently
been used to gain insight into different mechanisms such as
the molecular neurotransmitter signaling pathway \citep{nueroTrans}, the T cell receptor signaling pathway
\citep{julio_compBio_2007}, the signaling network for the long-term survival of cytotoxic T
lymphocytes in humans \citep{albert_plosBio_06},  and the abscisic acid signaling pathway \citep{albert_pnas_2008}.

Boolean models require less detailed information about the system to be modeled, so they can be used
in cases where quantitative information is missing. They are also useful if qualitative predictions from
the model are desired, such as whether a T cell becomes pro- or anti-inflammatory. Finally, Boolean models
are very intuitive compared to models based on differential equations or other more sophisticated
formalisms. It is also easier to explore their dynamics, at least for reasonably small models. On the other hand,
an important disadvantage of Boolean models, and algebraic models in general, is that there are very
few theoretical tools available for their construction. Typically, Boolean models are built by translating information
from the literature into logical statements about the interactions of the different molecular species involved in
the network. In many cases, the biological information about a particular network node might not be sufficient, however,
to construct a logical function governing regulation. 

In the case of a continuous model, the remedy would be to
insert a differential equation of specified form, \eg mass action kinetics, with unspecified parameters. If
experimental time course data are available one can then use one of several parameter estimation methods to
determine those unspecified model parameters so that the model fits the given data. Data fit is determined by
model simulation, using numerical integration of the equations in the model. The software package described
in this paper addresses the need for a discrete analog of this process.

In the case of missing information about a particular node in the network to be modeled one can insert a general
Boolean function, maybe of a specified type, \eg a nested canalyzing function. This is most easily done by
viewing the Boolean function as a general polynomial, with undetermined (0/1) coefficients. If experimental
time course data for the network is available, then one can use one of several existing inference methods
to estimate a function that will result in a model that fits the data. This function in addition satisfies a specified
optimality criterion, similar to the optimality criterion for the fitting of continuous parameters. This process
might be considered the discrete analog of parameter estimation. 

In this paper we describe a software package, \emph{Polynome}, which can be used for this purpose. The package
integrates several existing algorithms for parameter estimation and model simulation. Space limitations
do not allow a detailed self-contained description of each of the algorithms, most of whom have already appeared elsewhere. 
But enough detail is given so that the potential user understands the capabilities and limitations of the package.
We conclude the paper with an example application of the package to the well-known \emph{lac} operon,
the network that regulates lactose metabolism \emph{E. coli} . We use data generated from a Boolean model
of this network to illustrate software performance. 

\section{Architecture}
In this section we introduce the architecture of the software package \emph{Polynome} which
integrates algorithms that perform discrete parameter estimation, or system
identification, and simulation. A web interface of the software is available at

\begin{center}
\url{http://polymath.vbi.vt.edu/polynome/}.
\end{center}

The algorithms underlying the software represent Boolean networks as
time discrete dynamical systems as follows. Let $k = \{0,1\}$ be the field with two elements and
arithmetic modulo 2. A Boolean network in $n$ variables is a function
$$
f = (f_1,\ldots ,f_n): k^n \longrightarrow k^n,
$$
with $f_i\in k[x_1,\ldots , x_n]$. It is easy to see that any Boolean function can be represented as a polynomial
with coefficients in $k$. Furthermore, this polynomial can be chosen so that the variables appear only to the first power.
Two directed graphs are associated to this function. The \emph{wiring diagram} has as nodes the variables, and there
is a directed edge $i\rightarrow j$ if $x_i$ appears in $f_j$. The \emph{state space} of $f$ has as nodes all $2^n$ binary
strings in $k^n$. There is a directed arrow $\mathbf a\rightarrow \mathbf b$ if $f(\mathbf a ) = \mathbf b$.

There are two stochastic versions of Boolean networks that are relevant here. The first is update-stochastic networks.
Here, rather than updating the variables synchronously, they are updated asynchronously, using a randomly chosen
update order. Update-stochastic Boolean networks have been shown to capture interesting aspects of biological
networks \citep{sontag}
and stochastic sequential update is used in the most general form of the logical models introduced in \citep{LM_thomas}.
The second kind, function-stochastic Boolean networks are obtained by assigning a family of Boolean functions
to each node, together with a probability distribution on the family. At each update, a function from this family is
chosen at random. The software described here implements update stochastic networks as a subclass of
function-stochastic networks by including for each function also the identify function. Whenever the identity function
is chosen for an update, the corresponding variable is delayed, resulting in a sequential update.

\begin{figure}[!ht]
  \centering
\includegraphics[width=6.4in]{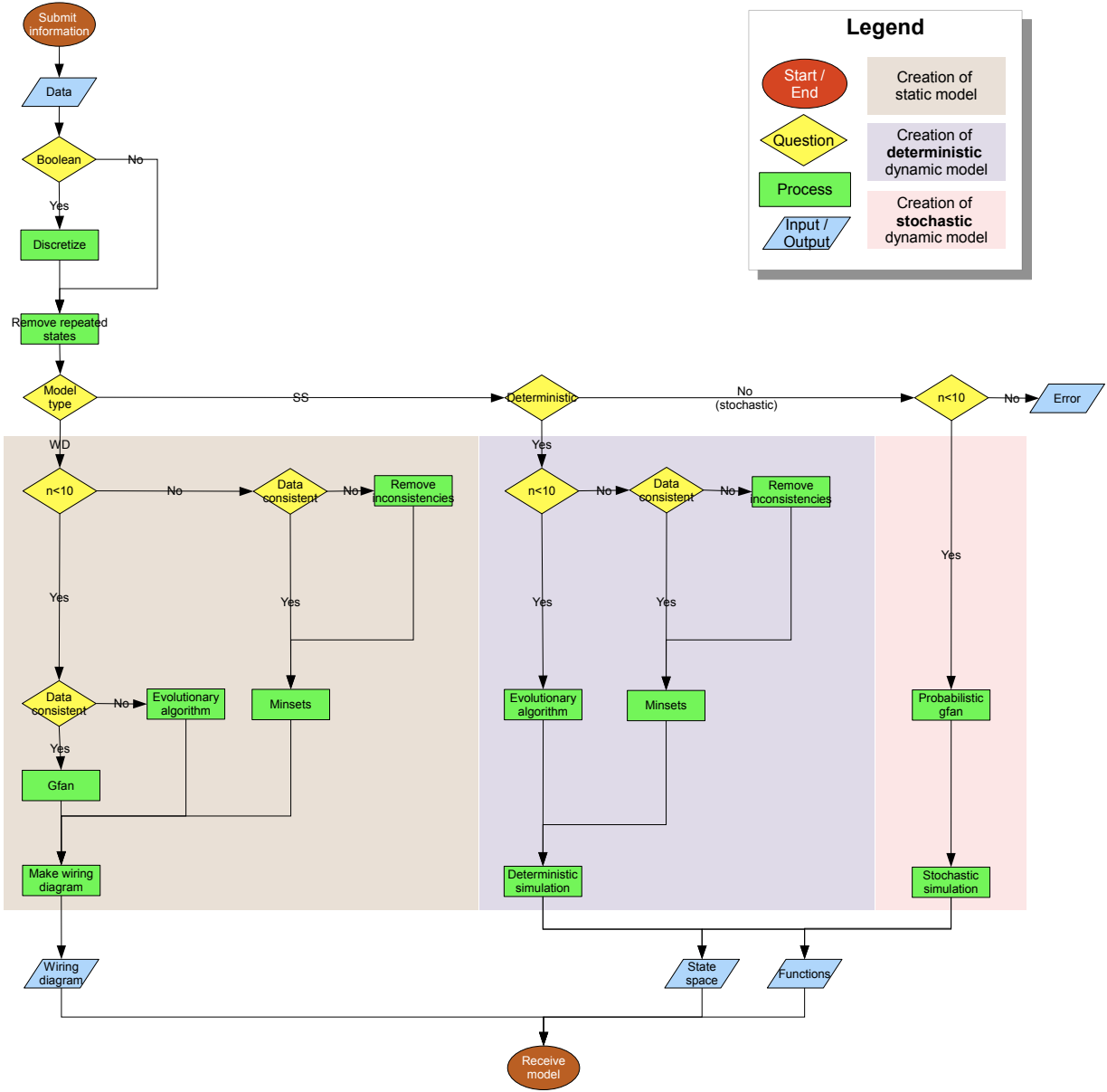}
  \caption{Flow chart of the software package Polynome.}
\label{flow-chart}
\end{figure}

Figure \ref{flow-chart} shows a flow chart of the software architecture.

The \textbf{input} consists of two parts:
\begin{enumerate}
\item
time course data, either continuous or Boolean (mandatory input);
\item
A subset of the $f_i$ (optional input).
\end{enumerate}
If the input consists of continuous data, then the software Booleanizes the
data first. The data need to be provided as a matrix with columns corresponding
to nodes and rows correspond to experimental data points, whereas the functions are input as a list.

There are several \textbf{output} options:
\begin{enumerate}
\item
A wiring diagram only,
showing the dependency relations between the variables of the network;
\item
a deterministic Boolean network model, which either fits the data exactly
or which optimizes between model complexity and data fit, and which
can be simulated either deterministically or stochastically;
\item
a stochastic Boolean network model.
\end{enumerate}

The simulator has several capabilities. It can:
\begin{enumerate}
\item
simulate a deterministic Boolean network and output the wiring diagram and/or the state space;
\item
simulate a deterministic Boolean network using random sequential updates of the variables and output
the state space with transition probabilities on the edges;
\item
simulate a function-stochastic Boolean network and output the state space with transition probabilities on the edges.
\end{enumerate}

The first step is to preprocess the data: Booleanize them if necessary (see Section \ref{sec:discretization}) and remove any states (rows) which occur more than once.  As there are many models which may fit a given data set, the set of possible models is typically very large, even with the minimality restriction.  We offer three ways to search the model space:
\begin{itemize}
    \item minimal-model sampling, based on the Gr\"obner fan sampling method (Algorithm \ref{dep-gr-reveng})
    \item minimal-model estimation, based on the method for noisy data (Algorithm \ref{EA1})
    \item minimal-model selection, based on the minimal-sets algorithm (Algorithm \ref{minsetsalg}).
\end{itemize}

For small networks ($n\leq 10$), the model space can be explored using one of two methods. Algorithm \ref{dep-gr-reveng} is used to sample the subspace of minimal models and returns a set of weighted functions per node (for stochastic models) or a set of weighted inputs per node (for static models - dynamics not desired by the user).  Algorithm \ref{EA1} is used to estimate the minimal Boolean networks in the model space when inconsistent data are provided or a deterministic model is desired.  What is returned is a polynomial dynamical system (PDS) that provides a best approximate data fit and is not overly complicated.  For moderate to large networks ($n>10$), the model space becomes too large to explore.  So Algorithm \ref{minsetsalg} is used to identify a subset of essential variables, and only models involving those variables are considered subsequently.  This algorithm returns either a minimal PDS (for large deterministic models) or a minimal wiring diagram (for static models - dynamics not desired by the user).  Note that Algorithms \ref{minsetsalg} and \ref{dep-gr-reveng} return PDSs that fit the data exactly.

Once a PDS has been identified, its dynamics can be simulated with one of the following modules: deterministic or stochastic simulation.  Given a PDS, its wiring diagram is  constructed using the GraphViz \texttt{dot} program.

\section{Data discretization}
\label{sec:discretization}

Discretization of continuous experimental data into finitely many discrete states is important for inferring gene regulatory networks from experimental data. Discretization of experimental data has been discussed
extensively, \eg  \citep{dim-discr-soft}.
The following definition of discretization is due to \citet{hartemink}.

\begin{definition}\label{discr-def}
A discretization of a real-valued vector $\mathbf{v}=(v_1,\dots,v_N)$ is an integer-valued vector $\mathbf{d}=(d_1,\dots,d_N)$ with the following properties:
\begin{enumerate}
	\item Each element of $\mathbf{d}$ is in the set ${0,1,\dots,D-1}$ for some (usually small) positive integer $D$, called the degree of the discretization.
	\item For all $1\leq i,j\leq N$, we have $d_i\leq d_j$ if and only if $v_i\leq v_j$.
\end{enumerate}
\end{definition}

Without loss of generality, assume that $\mathbf{v}$ is sorted, \ie for all $i < j$, $v_i\leq v_j$. Spanning discretizations of degree $D$ are a special case that we consider here. They are defined in \citep{hartemink} as discretizations that satisfy the additional property that the smallest element of $\mathbf{d}$ is equal to 0 and that the largest element of $\mathbf{d}$ is equal to $D-1$.
The translation from continuous to discrete data is crucial in preserving the variable dependencies and thus has a significant impact on the performance of the network inference algorithms. While there is a large selection of discretization methods available which cluster data points, many of them are not directly applicable in the network inference context or are not suitable. One important limitation is typically that the number of available data points is very small, typically consisting of less than 10 time points. We apply a newly developed method, based on graph theory, especially designed for short time series data \citep{dim-discr-soft}. Novel aspects are incorporation of an information-theoretic criterion and a criterion to determine the optimal number of values. While the method can be used on other types of data, the motivation for its development was the need for a discretization algorithm for several short multivariate time courses of heterogeneous data, such as transcript, protein, and metabolite concentration measurements. Furthermore, the method has been demonstrated to preserve the dynamic features of the time courses, as well as to be robust to noise in the experimental data.

The method begins by constructing a complete graph in which the vertices are the time points and the edge weights are the Euclidean distances between two vertices. Edges are deleted consecutively starting with the one of highest weight until the graph is disconnected. The process continues until one of the several stop criteria are met. The goal is to minimize the average internal distance of the components and maximize the distance between components.
In the current work we have limited the number of states to 2, \ie the data are Booleanized. The next version of \emph{Polynome} will be capable of 
handling parameter estimation for larger numbers of states. (The only parameter estimation algorithm that is currently not capable of handling something
other than binary states is REACT. A multi-state version is in preparation. However, in order to obtain useful performance, the computations will have to 
be performed in parallel on a multi-processor machine.) While Booleanization is a rather drastic transformation of the data and in many cases loses
valuable information in the data, it can still derive useful information from experimental data, as is shown in \cite{EA1}. There, the authors use transcript data
from a gene regulatory network in yeast used to compare different reverse-engineering methods \cite{dibernardo}. It is shown that the performance of REACT with a Booleanization of the data compares very favorably to the other methods tested. 

\begin{example}
Suppose that vector $\mathbf{v} = (1, 2, 7, 9, 10, 11)$ is to be discretized. We start by constructing the complete weighted graph based on $\mathbf{v}$.
\begin{figure}
\begin{center}
\includegraphics [width=3in]{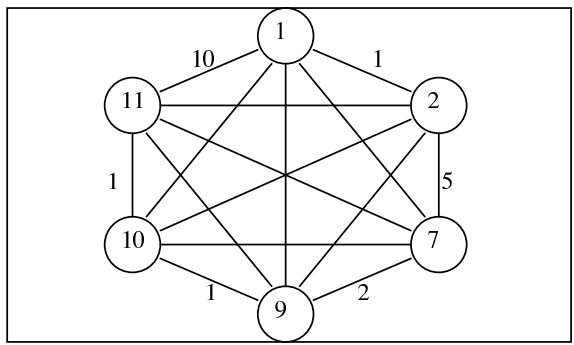}
\caption{The complete weighted graph constructed from vector entries 1, 2, 7, 9, 10, 11. Only the edge weights of the outer edges are given.}\label{graph}
\end{center}
\end{figure}
Eight edges with weights 10, 9, 9, 8, 8, 7, 6, and 5, respectively, have to be deleted to disconnect the graph into two components: one containing vertices 1 and 2 and another having vertices 7, 9, 10, and 11; this is the first iteration.
Having disconnected the graph, the next task is to determine if the obtained degree of discretization is sufficient; if not, the components need to be further disconnected in a similar manner to obtain a finer discretization.
\end{example}

A commonly occurring phenomenon, when discretizing time courses, is that the resulting data are inconsistent with a deterministic process. This happens
because a given state can transition to two different states at different times. So, when a deterministic model is desired, these inconsistencies have
to be removed. A common cause of such inconsistencies is small variations among consecutive time points, so that these get discretized into the
same state. Eventually, there is sufficient change in the data so that a later discrete state becomes different again. This situation is dealt with by
removing all but one instance of the repeated state.  This essentially amounts to a local adjustment of time scale. Since time is not represented 
explicitly in discrete models, this is permissible. In the case of a given state transitioning to two different states in two different time courses, we
remove the state in question, disconnecting the two time courses into four shorter ones. 
We further assume that there are no missing (unmeasured) time points. If new data points are included, then the parameter estimation process
has to be restarted at the beginning. 

\section{Parameter estimation}

\subsection{The minimal-sets algorithm}
Inferring the wiring diagram of a gene regulatory network has received lots of attention and many different methods
in different contexts have been developed to address this problem.
Using methods from computational algebra and algebraic geometry,
we have developed an algorithm that first finds all possible minimal wiring diagrams of a gene regulatory network, and
then chooses a particular model using different selection methods.
Here a diagram is \emph{minimal} if all specified interactions in the wiring diagram are necessary to have a function that
interpolates the data.

For a fixed gene, say $x_j$, we need to identify the  minimal sets of genes which could be used as inputs to $x_j$.
Let $\{({\mathbf s}_1,t_1),\ldots ,({\mathbf s}_m, t_m)\}$
be the stimuli-response data for the gene $x_j$, where ${\mathbf s}_i\in k^n, t_i\in k$. We need to find all minimal subsets $F\subset\{1, \ldots ,n\}$ such
that there exists a polynomial function $f\in k[\{x_i\mid i \in
F\}]$ with $f({\mathbf s}_i)=t_i$ and there is no such polynomial on any proper subset of $F$.
The main idea of the algorithm is the following: \\
For any two stimuli ${\mathbf s}_a$ and ${\mathbf s}_b$ such that $t_a \neq t_b$, identify
all coordinates $i$ such that ${\mathbf s}_{ai} \neq {\mathbf s}_{bi}$. If none of these coordinates was picked previously, pick one of them. Once all possible pairs are considered, the set of variables corresponding to the chosen coordinates can be used to generate the required function. This can be done using computational algebra algorithms, as described in \citep{LS}.

Using tools from computational algebra, the steps above can be
encoded as an algorithm that performs operations on a \emph{monomial ideal} $M$ with one generator for each
pair of stimuli encoding their mismatch, and each minimal prime of $M$
is a minimal set. See \citep{JLSS, SJLS}.

\begin{algorithm}[!ht]
\label{minsetsalg}
\caption{Minimal-Sets Algorithm}

\dontprintsemicolon
 \SetLine

\SetKwInOut{Input}{Input}
\SetKwInOut{Output}{Output}

\Input{ $\{({\mathbf s}_1,t_1),\ldots ,({\mathbf s}_m, t_m)\}$,
with ${\mathbf s}_i\in k^n, t_i\in k$}

\Output{All minimal subsets $F\subset\{1, \ldots ,n\}$ such
that there exists a polynomial function $f\in k[\{x_i\mid i \in F\}]$ with $f({\mathbf s}_i)=t_i$.}

\BlankLine

\Begin{
\begin{enumerate}[Step 1.]
\item Compute the ideal $M$.
\item Compute a primary decomposition of $M$.
\item Output the generating sets of all minimal primes of
$M$.
\end{enumerate}
}
\end{algorithm}

It is clear that there will usually be many possible wiring diagrams and
selecting a single model could be done only based on further assumptions,
such as knowledge from the literature about some of the interactions,
or the network being sparse.
For the purpose of building a web-based application, we employ the $(S_1,T_1)$ ranking scheme described in \citep{JLSS} and used in \citep{SJLS} to select highest-scoring minimal sets.  The scheme ranks sets according to size (smaller is better) and frequency of occurrence of the variables (higher is better).  Given a list of highest-scoring sets of equal rank, we choose the first one in the list.


\subsection{Parameter estimation using the Gr\"{o}bner fan of an
ideal } \label{sec:par_est_groebner_fan}

Typically there are many models that fit a time course of experimental data and often there is insufficient information to select one of them. Even restricting the model space to minimal models leaves multiple possible models. The reason is that different orderings of the polynomial terms (monomials) can give rise to different polynomial models, since the algorithm uses such an order for multivariate polynomial division.
Algorithm \ref{dep-gr-reveng} generates all Boolean network models fitting the data and, if needed, extracts the corresponding wiring diagram. The method is based on a combinatorial structure known as the \emph{Gr\"{o}bner fan} of a polynomial ideal \citep{morarobb, sturmfels}, which is a polyhedral complex of
cones in which every point encodes a monomial ordering. The cones are in bijective correspondence with the distinct Gr\"{o}bner bases of an ideal. (To be precise, the correspondence is to the marked reduced Gr\"{o}bner bases of the ideal). Therefore, it is sufficient to select exactly one monomial ordering per cone and, ignoring the rest of the orderings, this still guarantees that all distinct minimal models are generated. In addition, the relative number of monomial orderings under which a particular PDS model is generated gives us insight into the likelihood that the model is a good representation of the system. The sizes of the Gr\"{o}bner cones can be computed for small Gr\"{o}bner fans using the method of \citep{yoshida} or sampling uniformly a large number of points. The monomial orderings used in generating the models are selected through random sampling of the corresponding Gr\"{o}bner fan of the ideal of points. If the number of points is sufficiently large, their distribution approximately reflects the relative size of the Gr\"{o}bner cones. The number of points is determined using a $t$-test hypothesis testing for proportion.

Steps 1--5 of Algorithm \ref{dep-gr-reveng} are used for parameter estimation of a stochastic model of a system. If the wiring diagram is also required, Step 6 is performed as well. If the Gr\"{o}bner fan is too large to compute, Step 3 is replaced by a large random sample of points from the Gr\"{o}bner fan which reflects the relative sizes of the Gr\"{o}bner cones.

\begin{algorithm}[!ht]
\label{dep-gr-reveng}
\caption{Parameter sampling using the relative sizes of the Gr\"{o}bner cones.}

\dontprintsemicolon
 \SetLine

\SetKwInOut{Input}{Input}
\SetKwInOut{Output}{Output}

\Input{A discrete time course of a network on $n$ nodes $x_1,\ldots, x_n$:
 $S=\{(s_{11},\dots,s_{n1}),\dots,(s_{1m},\dots,s_{nm})\}\subseteq k^n$.}

\Output{A list of length $n^2$: $\{p_{ij}~|~i,j=1,\ldots,n\}\subset\left[0,1\right]$ where $p_{ij}$ is the strength with which $x_i$ affects $x_j$.}

\BlankLine

\Begin{
\parbox{5.5in}{
\begin{enumerate}[Step 1.]
\item Compute a particular Boolean network $F_0:k^n\rightarrow k^n$ that fits $S$.

\item Compute the ideal $I$ of polynomials that vanish on $S$.

\item Compute the Gr\"{o}bner fan $\mathcal{G}$ of the ideal $I$ and the relative sizes of its cones, $c_1,\ldots,c_s$ (with $c_1+\cdots+c_s$=1).

\item Select one (any) monomial ordering from each cone, $\prec_1,\ldots,\prec_s$. For each $i=1,\ldots,s$, reduce $F_0$ modulo $I$ using a Gr\"{o}bner basis computed with respect to $\prec_i$. Let the reduced PDSs be $F_1,\ldots,F_s$ and form the 2-tuples $(F_i,c_i)$.

\item If $F_i=F_j$ for some $i,j=1,\ldots,s$, then ``merge'' the two 2-tuples into $(F_i,c_i+c_j)$. After re-indexing, obtain the set $\{(F_1,c_1),\ldots,(F_t,c_t)\}$ where $t\leq s$ and $F_i\neq F_j$ for all $i\neq j$.

\item For every $i,j=1,\ldots,n$, find all Boolean networks $F_{l_1},\ldots,F_{l_d}$ whose dependency graph contains a directed edge $x_i\mapsto x_j$ and let $p_{ij}=c_{l_1}+\cdots+c_{l_d}$. If no Boolean network contains the edge $x_i\mapsto x_j$ for some $i$ and $j$, set $p_{ij}=0$.
\end{enumerate}
}
}
\end{algorithm}




\subsection{Parameter Estimation in the Presence of Data Noise}

Since experimental data are typically noisy, due to
measurement error or intrinsic biological noise, robustness
to noise of inference methods is desirable. Therefore, in order to avoid over-fitting,
in \citep{EA1} a method for parameter estimation is described, based
on the premise that the input data may contain \textbf{noise}. In this genetic algorithm
based method,  a Boolean network model is inferred which is optimized with respect to both data fit and
model complexity; an optimal Boolean model can also be constructed
when prior knowledge about the network structure is included.\\

Consider a network of $n$ nodes. The main elements of the genetic algorithm (GA) are
\begin{itemize}
\item Chromosomes of GA: Polynomial models, where a polynomial model is given as a system of $n$ polynomial functions $F_1, F_2,\dots,F_n$.
\item Genes of GA: Polynomial functions (hence there are $n$ different classes of genes).
\item Each chromosome (polynomial model) in the GA contains $n$ genes ($n$ polynomial functions $F_1, F_2,\dots,F_n$).
\end{itemize}

Algorithm \ref{EA1} summarizes the key features in this procedure.

\begin{algorithm}[!ht]
\label{EA1}
\caption{Parameter estimation in the presence of data noise}

\dontprintsemicolon
 \SetLine

\SetKwInOut{Input}{Input}
\SetKwInOut{Output}{Output}

\Input{Boolean time courses (wildtype and/or knockout biological data).
(Optional input: seeding polynomial models (seeding chromosomes) and/or prior knowledge of network structure)}

\Output{Polynomial models with the best fitness scores.}

\BlankLine
\Begin{
\parbox{5.5in}{
\begin{enumerate}[Step 1.]

\item  If no \emph{seed} is provided, generate random polynomial
models until the pool size is reached. Otherwise, use input models and
complete pool size with randomly generated models;

\item Evaluate and sort models according to their fitness;

\item Subdivide \textit{population} into $n$ \textit{sub populations},
each corresponding to one of the $n$ coordinate functions;

\item Select polynomials from each subpopulation corresponding to each of
the coordinate functions and assemble them into new polynomial
models to form the generation's \textit{offspring};

\item Mutate the assembled candidate models  with inverse probability
to the fitness of the generation, \ie fewer mutations will take place in a generation
that shows higher fitness scores;

\item \textit{Clone} polynomial models selected to be preserved in the
next generation;

\item Build new population from candidate models, \textit{cloned} models
and randomly generated models;

\item Repeat Steps 2 through 7 until either the specified number of
generations is reached or the fitness score has not improved for a pre-determined
number of generations.
\end{enumerate}
}
}
\end{algorithm}

Some comments about the implementation of this algorithm in \emph{Polynome}
are in order. 

{\bf{Model fitness.}} The fitness function used in the GA is a multi-objective function
incorporating the different fitness criteria of each Boolean coordinate function for each 
node in the network as well as the fitness of the fully assembled Boolean models.
The different criteria include: data fit, model complexity, and consistency
with prior knowledge about the network structure.

{\bf{Parameters.}} The algorithm is controlled by a set of parameters that 
specify properties such as gene pool size and an upper bound for the maximum number
of generations to run the GA. In the current version of the software, a default selection of parameters
is made based on a preselected maximum network size. 
In future versions of the algorithm, 
the user will be able to control such parameters for an \textit{ad-hoc} selection
based on the specific network analyzed by the user.

{\bf{Termination criteria.}} Common concerns about genetic algorithms are the selection of  termination criteria, how to avoid
local minima, or to the computational resources required to run the algorithm for sufficiently many generations. 
Our stopping criterion is based on two parameters: a parameter
that controls the maximum number of generations to be run and the maximum number of generations 
after which the algorithm is terminated if the fitness score has not improved. Both of these parameters have been selected
based on the current limit on network size and the need of the web service version to provide output within a short
period of time. It is important to mention that this limitation generally leads to models of lower quality than one would
obtain from a free-standing version of the software.

{\bf{Output.} }The algorithm described in \cite{EA1} provides as output all models with highest fitness score. Due to interface
limitations, the implementation in \emph{Polynome} outputs only ten of these models. Future versions
of \emph{Polynome} will allow the user to request more models. As with all evolutionary algorithms, it is not possible
to prove convergence results, guarantee that the results are not just local optima, or guarantee that the search results are robust under
repetition. These are all drawbacks of this method over the other methods available in \emph{Polynome}. It is also possible
that two models with the same score are very different from each other. On the other
hand, this algorithm is the only one that can produce models which do not have to fit the given data exactly, so they are more
robust to noise. 

\section{Simulation}
For the identified network, fixed points and limit cycles are calculated and the
phase space and wiring diagram are generated.  The analog of the graph of a
function in the continuous case, \eg solution of a system of differential
equations, is the phase space for a discrete model. It visualizes the dynamics
of the network, namely the fixed points and oscillatory cycles.
 In a deterministic network,
each state has out-degree exactly $1$, in a function stochastic network as
generated by \ref{sec:par_est_groebner_fan}, the
out-degree can be higher.
In a
stochastic network fixed points have a stability that is calculated from the
probabilities of the update functions. The stability indicates how likely it
is to remain in this state, so stability 1 corresponds to a true fixed point.

%
%
For a function stochastic system a synchronous update is used, but for a
deterministic network it is possible to use a sequential update order instead,
in case more biological information about the network is available, such as the order in
which certain molecular processes take place.
 If the user wants to use sequential update without
providing an update order, then the software uses stochastic sequential update, that is,
at each update an update order is chosen at random. As pointed out in the introduction,
sequential update has been shown to be biologically more realistic (see, e.g., \cite{sontag}), which is the reason
why we are providing this simulation capability. However, it is easy to see that different
update orders result in different dynamics, so that it is possible that a deterministic system
that was chosen to fit a given set of experimental data will not do so any longer when
simulated sequentially. 

%
%

\section{An example: the \emph{Lac} operon}

We demonstrate the key features of the software with an example. For simplicity, we choose an existing Boolean network
model in order to be able to compare the output to the ``true" network.
We consider a Boolean model \citep{SV} for lactose metabolism in the context of the \emph{lac} operon for the following two examples.  Let $f$ be the 9-node Boolean model in \citep{SV} in terms of the variables $(M,P,B,C,R,A,A_l,L,L_l)$ and the parameters $(L_e,G_e)$ (see the original manuscript for an introduction to the \emph{lac} systems and a description of the model).  For simplicity, we rename the variables as $(x_1,\ldots,x_9)$ and the parameters as $(x_{10},x_{11})$, and write the Boolean functions as in Table \ref{bool},
where $f_i$ represents the Boolean function associated to variable $x_i$ and $\sim$ is the logical NOT operator, $*$ AND, and $+$ OR.

\begin{table}[!ht]
    \centering
    \begin{tabular}{|l|l|}
        \hline
        $\begin{array}{l}
            f_M=(\sim  R)* C \\
            f_P=M\\
            f_B=M\\
            f_C=\sim  G_e\\
            f_R=(\sim  A) * (\sim  A_l) \\
            f_A=L* B\\
            f_{A_l}=A+ L+ L_l\\
            f_L=(\sim  G_e)* (P* L_e)\\
            f_{L_l}=(\sim  G_e)*(L+ L_e)
        \end{array}$
        &
        $\begin{array}{l}
            f_1 = (\sim x_5) * x_4\\
            f_2 = x_1\\
            f_3 = x_1\\
            f_4 = \sim x_{11}\\
            f_5 = (\sim x_6) * (\sim x_7)\\
            f_6 = x_8 * x_3\\
            f_7 = x_6 + x_8 + x_9\\
            f_8 = (\sim x_{11}) * (x_2 * x_{10})\\
            f_9 = (\sim x_{11}) * (x_8 + x_{10})
        \end{array}$ \\
        \hline
    \end{tabular}
    \caption{Left panel: Boolean functions in the original variables.  Right panel: Boolean functions in the indeterminates $x_1,\ldots,x_{11}$.}
    \label{bool}
\end{table}

Note that functions in \emph{Boolean form}, with binary operations $*$ and $+$ and unary operation $\sim$ as defined above, can be translated to \emph{polynomial form}, with field operations $+$ and $\times$ (also written as $*$ in nonformatted text) via the mapping
\begin{center}
\begin{tabular}{|c|c|}
  \hline
  Boolean form & Polynomial form \\
  \hline
  $\sim x$ & $x+1$ \\
  $x*y$ & $xy$ \\
  $x+y$ & $xy+x+y$ \\
  \hline
\end{tabular}
\end{center}

In the following example, we consider the case where we wish to identify some of the functions in a partially known network.

\begin{example}
In the Boolean model in Table \ref{bool}, the functions for $M, P, B, C,$ and $R$ are straight-forward from a biological perspective. However, this is not the case for the functions for lactose ($L, L_l$) and allolactose ($A, A_l$). There is only one combination of values for extracellular glucose ($G_e$) and lactose ($L_e$) for which the operon is ON; \ie $L_e=1, G_e=0$. Setting the parameters to these values produces a single fixed point $(1,1,1,1,0,1,1,1,1)$ in the above Boolean model.  Further it limits what the functions for lactose ($L, L_l$) could be, namely $f_L = P$ and $f_{L_l} = 1$. This leaves $f_A$ and $f_{A_l}$ open for investigation.

For demonstrative purposes we chose the following data sets, which represent immediate initiation ($C=1$) of the \textit{lac} operon when applicable,
to use in the parameter estimation step.

\begin{tabular}{|l|l|l|}
  \hline
Operon is OFF & Operon is ON\\
\hline
$\begin{array}[t]{ccccccccc}
0&0&0&1&0&0&0&0&0\\
1&0&0&1&1&0&0&0&1\\
0&1&1&1&1&0&1&0&1\\
0&0&0&1&0&0&1&1&1\\
1&0&0&1&0&0&1&0&1\\
1&1&1&1&0&0&1&0&1\\
1&1&1&1&0&0&1&1&1\\
1&1&1&1&0&1&1&1&1\\
1&1&1&1&0&1&1&1&1
\end{array}$
&
$\begin{array}[t]{ccccccccc}
1&1&1&1&1&1&1&1&1\\
0&1&1&1&0&1&1&1&1\\
1&0&0&1&0&1&1&1&1\\
1&1&1&1&0&0&1&0&1
\end{array}$ \\
\hline
Transcription of \textit{lac} genes & Repression of operon \\
\hline
$\begin{array}[t]{ccccccccc}
1&0&0&1&0&0&0&0&0\\
1&1&1&1&1&0&0&0&1\\
0&1&1&1&1&0&1&1&1\\
0&0&0&1&0&1&1&1&1\\
1&0&0&1&0&0&1&0&1
\end{array}$
&
$\begin{array}[t]{ccccccccc}
0&0&0&1&1&0&0&0&0\\
0&0&0&1&1&0&0&0&1\\
0&0&0&1&1&0&1&0&1\\
0&0&0&1&0&0&1&0&1\\
1&0&0&1&0&0&1&0&1
\end{array}$ \\
\hline
Initiation by $A, A_l$ & Initiation by $L, L_l$ \\
\hline
$\begin{array}[t]{ccccccccc}
0&0&0&1&0&1&1&0&0\\
1&0&0&1&0&0&1&0&1
\end{array}$
&
$\begin{array}[t]{ccccccccc}
0&0&0&1&0&0&0&1&1\\
1&0&0&1&1&0&1&0&1\\
0&1&1&1&0&0&1&0&1\\
1&0&0&1&0&0&1&1&1\\
1&1&1&1&0&0&1&0&1
\end{array}$ \\
  \hline
\end{tabular}

Given the above data, Algorithm \ref{dep-gr-reveng} is used to identify the connections (see Figure \ref{wd-bool}), as well as the functions in polynomial form for allolactose as follows:

\begin{figure}[!ht]
  \centering
  \includegraphics[width=5in]{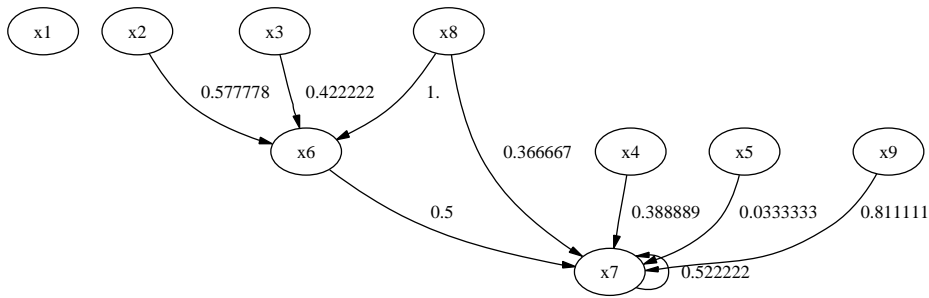}\\
\caption{Inferred wiring diagram for allolactose.}
  \label{wd-bool}
\end{figure}

\medskip
\hrule
$$\vdots$$
\begin{verbatim}
f6 = {
x3*x8   0.422222
x2*x8   0.577778
}
f7 = {
x4*x8+x5*x7+x8+x4*x5+x7   0.0111111
x7*x8+x8+x4*x7+x7+x4      0.0333333
x6*x9+x6+x9               0.111111
x7*x9+x7+x9               0.322222
x4*x6+x6+x9               0.122222
x6*x8+x6+x9               0.177778
x4*x8+x8+x4*x6+x6+x4      0.0222222
x4*x8+x8+x4*x7+x7+x4      0.0444444
x4*x7+x7+x9               0.0666667
x7*x8+x5*x7+x8+x4*x5+x7   0.0111111
x6*x8+x4*x8+x8+x6+x4      0.0444444
x7*x8+x6*x8+x8+x6+x4      0.0222222
x5*x7+x5*x9+x7+x9+x4      0.0111111
}
\end{verbatim}
$$\vdots$$
\hrule
\medskip
The figure represents only \emph{part} of the full wiring diagram; we include only the subgraph of edges incident to nodes 6 and 7 for simplicity and include the isolated node $x1$ to emphasize that $x_1$ is not an input of either $x_6$ or $x_7$. The edge weights are values between 0 and 1 and can be interpreted
as relative likelihood of interaction or interaction strength. That is, if the weight on the edge from node $i$ to node $j$ is $p$, this means
that the relative likelihood or strength with which $i$ affects $j$ is $p$.

In the table of functions, again we only display the portion of the output that is relevant in this example.  We find 2 possibilities for $f_6$, one which matches the original function, and 13 possibilities for $f_7$.  Note that the original for $f_7$ can be written as $$x_8f_{7,3} + f_{7,3} + x_8$$ in polynomial form, or $$x_8 + f_{7,3}$$ in Boolean form, where $f_{7,3}$ is the third element in the function list \texttt{f7}.  The weights associated to each function can be interpreted similarly to the weights on the edges in the wiring diagram.  Since we have multiple possibilities for each function, we can produce a stochastic simulation of the model.  This yields the following dynamics:
\medskip
\hrule
\begin{verbatim}
Number of components 1
Number of fixed points 2
Fixed point, component size, stability
(1 1 1 1 0 1 1 0 1), 512, 0.08
(1 1 1 1 0 1 1 1 1), 512, 1.00
\end{verbatim}
\hrule
\medskip
The first fixed point has very small stability and is therefore not reliable as a steady state.  However, the second one, which corresponds to the unique fixed point in the original Boolean model, has stability 1 indicating that it is a true steady state.

In Figure \ref{ss-bool} we show the portion of the state space.  Since two of the functions are constant, we remove them for simplicity.  We note that there is a greater probability (0.92) of encountering the state in which all molecules are present, as opposed to the low probability (0.08) of encountering the state in which all molecules are present except for $A_l$, which is biologically infeasible, from the state (1,1,1,\emph{1},0,0,0,1,\emph{1})\footnote{Italicized coordinates where removed from the figure.}.
\begin{figure}[t]
    \centering
  \includegraphics[width=2in]{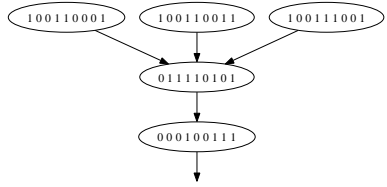}\\
  \caption{Portion of the state space of the Boolean model simulated stochastically.}
  \label{ss-bool}
\end{figure}

\end{example}

\begin{example}
Here we consider the case that we do not know any of the functions and aim to construct a deterministic dynamic model.  Given that the data are consistent, the default strategy is to produce a model using the minimal-sets algorithm.  However, for demonstrative purposes, we illustrate Algorithm \ref{EA1} which is reserved for inconsistent or noisy data.  Using the above data, we get the following model and its dynamics:
\medskip
\hrule
\begin{verbatim}
f1 = x1*x6*x7 + x1*x6 + x5 + 1
f2 = x1 + 0
f3 = x1 + x2 + x3*x4 + x4*x9 + x4 + 0
f4 = 1
f5 = x7 + 1
f6 = x2*x4*x8 + x2*x6 + x4 + x5*x7*x8 + x5*x8 + x6*x9 + x6 + x9 + 0
f7 = x1*x3*x6 + x6*x8 + x9 + 0
f8 = x3 + 0
f9 = 1
\end{verbatim}
\hrule
\begin{verbatim}
There are 7 components and 3 fixed point(s)
\end{verbatim}
$$\vdots$$
\begin{verbatim}
[ 0 0 1 1 1 1 0 1 1 ] lies in a component of size 6.
[ 1 1 0 1 0 0 1 0 1 ] lies in a component of size 4.
[ 1 1 0 1 0 1 1 0 1 ] lies in a component of size 124.
\end{verbatim}
\hrule
\medskip

It is important to mention that Algorithm \ref{EA1}  returns a list of the highest-scoring models (according to an internal fitness score). Hence in
some instances like in this example, more than one model is returned with the same high score:
\medskip
\hrule
\begin{verbatim}
f1 = x5 + 1
f2 = x1 + 0
f3 = x1 + x2 + x3*x4 + x4*x9 + x4 + 0
f4 = 1
f5 = x7*x9 + 1
f6 = x3*x8 + 0
f7 = x4 + 0
f8 = x2*x7 + x2 + x3*x7 + x5*x6*x8 + 0
f9 = 1
\end{verbatim}
\hrule
\begin{verbatim}
[ 1 1 0 1 0 0 1 0 1 ] lies in a component of size 256.
[ 1 1 1 1 0 1 1 1 1 ] lies in a component of size 256.
\end{verbatim}
\hrule
\medskip
Both models have the highest ranking, meaning that they both fit the given data well, though not exactly.  However, the second model has dynamics which resembles the original Boolean model: it only has fixed points (the first model has 4 nontrivial cycles) and one matches the unique fixed point of the Boolean model.
\end{example}

\section{Discussion}

We have presented a description of a software package to construct models of biological networks by fitting
Boolean network models to time course experimental data. The software is offered as a web application.
A detailed tutorial is available to help the user. Several other features will be incorporated in the next
release, including the ability of the user to specify that the software return nested canalyzing Boolean
functions, a particular type of Boolean function that was introduced by S. Kauffman and his collaborators
\citep{Kauff2}. The next release will also allow the user to infer polynomial dynamical systems with more
than two states. In addition, more useful graphical features will be introduced.

\section*{Acknowledgements}The authors are grateful to the Statistical and Applied Mathematical Sciences Institute (SAMSI),
the Center for Discrete Mathematics and Theoretical Computer Science (DIMACS), and the Mathematical Biosciences Institute (MBI)
for supporting the work reported in this paper. Dimitrova was partially supported by NSF/EPSCoR Award Nr. EPS-0447660,
and Garcia and Laubenbacher were partially supported by SAMSI. Stillman was partially supported by NSF Award Nr. DMS 08-10909.



\bibliographystyle{elsarticle-harv}

\end{document}